\documentclass[twocolumn,aps,prl,superscriptaddress]{revtex4-1}
\usepackage{amsmath}
\usepackage{bm}
\usepackage{graphicx}
\usepackage{mathrsfs}
\usepackage{multirow}
\usepackage[normalem]{ulem}
\usepackage{graphicx}
\usepackage{booktabs, ctable}
\usepackage{xcolor, color, framed}
\usepackage[colorlinks, linkcolor=red,anchorcolor=green,citecolor=blue]{hyperref}

\graphicspath{{./figs/}}

\begin{document}

\title{Quantum Localization Limit of Transport-Based Femtoscopy}

 \author{Jiaxing Zhao}
 \email{jzhao@itp.uni-frankfurt.de}
 \affiliation{Helmholtz Research Academy Hesse for FAIR (HFHF), GSI Helmholtz Center for Heavy Ion Physics, Campus Frankfurt, 60438 Frankfurt, Germany}
 \affiliation{Institut f\"ur Theoretische Physik, Johann Wolfgang Goethe-Universit\"at,Max-von-Laue-Straße 1, D-60438 Frankfurt am Main, Germany}

\author{Joerg Aichelin}
\email{aichelin@subatech.in2p3.fr}
\affiliation{SUBATECH UMR 6457 (IMT Atlantique,  Universit\'{e} de Nantes, IN2P3/CNRS), 4 Rue Alfred Kastler, F-44307 Nantes, France}
\affiliation{Frankfurt Institute for Advanced Studies, Ruth-Moufang-Strasse 1, 60438 Frankfurt am Main, Germany}

\author{Elena Bratkovskaya}
\email{E.Bratkovskaya@gsi.de}
\affiliation{GSI Helmholtzzentrum f\"{u}r Schwerionenforschung GmbH, Planckstrasse 1, 64291 Darmstadt, Germany}
\affiliation{Helmholtz Research Academy Hesse for FAIR (HFHF), GSI Helmholtz Center for Heavy Ion Physics, Campus Frankfurt, 60438 Frankfurt, Germany}
\affiliation{Institut f\"ur Theoretische Physik, Johann Wolfgang Goethe-Universit\"at,Max-von-Laue-Straße 1, D-60438 Frankfurt am Main, Germany}

\date{\today}

\begin{abstract}
Femtoscopic correlations in proton-proton and heavy ion collisions are commonly calculated using emission sources
from microscopic transport models, where emitted particles are
represented by classical phase-space points with simultaneously
specified positions and momenta. This point-emitter picture neglects
the finite phase-space localization required by quantum mechanics,
which can become relevant when the localization scale approaches the
femtoscopic source size, particularly in small collision systems.
We formulate a quantum treatment by replacing each transport
phase-space point with a minimum-uncertainty Gaussian phase-space
distribution. Finite localization then induces both spatial and
momentum smearing; in the presence of coordinate--momentum
correlations, the latter produces a nontrivial modification of the
emission source. Using proton--proton collisions at
$\sqrt{s}=7~\mathrm{TeV}$ simulated with the microscopic Parton--Hadron--String
Dynamics transport approach, we show that this effect significantly modifies the
proton-pair source and its momentum correlation.
Our results expose a quantum localization effect overlooked in
transport-based femtoscopy and establish the regime in which the
classical point-emitter approximation is valid.
\end{abstract}

\maketitle

Hadron--hadron interactions provide essential insight into the nonperturbative dynamics of quantum chromodynamics (QCD) and the structure of conventional and exotic hadrons. While interactions between stable hadrons can be constrained by scattering experiments, such measurements are challenging or impossible for short-lived particles, leaving many interactions involving hyperons and heavy-flavor hadrons poorly known. Two-particle femtoscopy offers a powerful alternative, as correlations at small relative momentum encode the final-state interaction and thereby provide access to the underlying hadron--hadron interaction
\cite{Koonin:1977fh,Lednicky:1981su,Pratt:1986cc,Lisa:2005dd,
Mihaylov:2018rva,Fabbietti:2020bfg,Ohnishi:2016elb,Kamiya:2021hdb,Torres-Rincon:2023qll,Mihaylov:2023ahn,Liu:2024uxn,Yan:2024aap,Encarnacion:2024jge,Mrowczynski:2025qys}.

High-statistics measurements in $pp$ collisions have enabled
precision studies of $p(K)\Lambda$~\cite{ALICE:2018ysd,ALICE:2023wjz},
$\Lambda\Lambda$~\cite{ALICE:2018ysd}, $p\Sigma$~\cite{ALICE:2025plu},
$p\Omega$~\cite{ALICE:2020mfd}, $\Lambda\Xi$~\cite{ALICE:2022uso},
$p(\pi,K)\Xi$~\cite{ALICE:2020mfd,ALICE:2025kma},
$p\phi$~\cite{ALICE:2021cpv}, $\pi(K)D$~\cite{ALICE:2024bhk}
interactions, and so on. In the Koonin--Pratt formalism~\cite{Koonin:1977fh},
the correlation function is given by
\begin{equation}
 C(k^*)=\int d^3r^*\,S({\bm r}^*|{\bm k}^*)
 \left|\Psi_{{\bm k}^*}^{(-)}({\bm r}^*)\right|^2 ,
 \label{eq:KP_intro}
\end{equation}
where $S({\bm r}^*|{\bm k}^*)$ is spatial distribution conditioned on observing the pair with momentum ${\bm k}^*$ and
$\Psi_{{\bm k}^*}^{(-)}({\bm r}^*)$ is the two-particle scattering wave
function. The superscript $^*$ is used to label quantities evaluated in the pair rest frame (PRF).
The extraction of the interaction therefore relies critically
on an accurate determination of the emission source, particularly in
small collision systems where its characteristic size is itself of
femtometer scale~\cite{ALICE:2023sjd,ALICE:2020ibs}.

Experimental femtoscopic analyses frequently employ the
Lednick\'y--Lyuboshits analytical approach~\cite{Lednicky:1981su}, in
which the emission source is commonly parametrized by a Gaussian
distribution characterized by an effective source radius. Realistic
particle production, resonance decays, rescattering, and collective
dynamics, however, generate non-Gaussian structures and
coordinate--momentum correlations~\cite{Wiedemann:1996ig,Kisiel:2006is,
ALICE:2023sjd}, making quantitative control of the emission source
essential for extracting hadronic interactions~\cite{Epelbaum:2025aan}.
Microscopic event generators provide a natural alternative by
constructing $S({\bm r}^*|{\bm k}^*)$ directly from the space--time
coordinates and momenta of emitted particles
~\cite{Li:2008qm,Kincses:2022eqq,Karpenko:2010te,Kisiel:2014upa,
Porfy:2025sev,Kincses:2025izu,Zhao:2026tpj,Zhao:2026bhh,
Ramirez:2026qfb}. Such sources can be combined with numerically
calculated two-body scattering wave functions, as implemented, for
example, in the Correlation Analysis Tool using the Schr\"odinger
equation (CATS)~\cite{Mihaylov:2018rva} and the Correlation Afterburner
(CRAB)~\cite{Pratt:1986cc}.

This microscopic approach nevertheless leaves an intrinsically quantum
aspect of the emission source unresolved. A fully quantum treatment
of the many-body collision dynamics is computationally prohibitive for
realistic high-energy reactions, and transport approaches instead
represent emitted particles by classical phase-space points with
simultaneously specified positions and momenta. In conventional
transport-based femtoscopy, these points are subsequently treated as
pointlike emitters, neglecting the finite phase-space localization
required by quantum mechanics. A spatially localized emitter
necessarily carries a finite momentum uncertainty. Quantum localization
can therefore modify not only the spatial source but, through
coordinate--momentum correlations, also the source associated with a
given relative momentum. This effect is expected to be particularly
relevant in small collision systems, where the source size can approach
the quantum localization scale and the classical point-emitter
approximation is no longer parametrically guaranteed.

In this Letter, we develop a quantum wave-packet formulation of
transport-based femtoscopy by associating each classical phase-space
point with a minimum-uncertainty Gaussian wave packet. The corresponding
Wigner formulation introduces finite uncertainties in both relative
coordinate and momentum. For a factorized Gaussian source, this effect
reduces to a simple broadening of the effective source radius, whereas
coordinate--momentum correlations in realistic transport sources lead
to an additional, momentum-dependent modification of the emission
source. Using the microscopic off-shell Parton--Hadron--String Dynamics
(PHSD) transport approach~\cite{Cassing:2008sv,Cassing:2009vt,Moreau:2019vhw},
which describes the dynamical evolution of interacting partonic and
hadronic degrees of freedom, we simulate $pp$ collisions at
$\sqrt{s}=7$~TeV and demonstrate sizable modifications of both the
proton-pair emission source and the resulting low-relative-momentum
femtoscopic correlation. Finally, we establish a general criterion for
the validity of the classical point-emitter approximation.

In event-generator-based femtoscopy, the space-time and momentum coordinates of the emitted particles are represented by
classical phase-space points. The two particles of a given pair generally undergo their last interactions at different space-time points, $x_i^\mu=(t_i,{\bm x}_i)$ and $p_i^\mu=(E_i,{\bm P}_i)$, with $i=1,2$, and in general $t_1\neq t_2$. The transport evolution itself is performed in the simulation frame (the laboratory frame for collision experiments), while the femtoscopic variables are evaluated in the PRF. The transformation to the PRF is performed only after freeze-out in order to construct the
relative variables entering the two-particle correlation function. For each pair, the Lorentz transform is done from the simulation frame to the PRF, $x_i^\mu\rightarrow x_i^{*\mu}
 =(t_i^*,{\bm x}_i^*)$, $p_i^\mu\rightarrow p_i^{*\mu}
 =(E_i^*,{\bm P}_i^*)$. Since the two freeze-out events are generally non-simultaneous in the PRF,
$t_1^*\neq t_2^*$. To construct the equal-time relative separation
entering the Koonin--Pratt formalism, the earlier emitted particle is
freely propagated to the emission time of the later particle. Defining $t_{\rm max}^*=\max(t_1^*,t_2^*)$, the equal-time positions are $\widetilde{\bm x}_i^*
 ={\bm x}_i^*+ {\bm v}_i^*
 \left(t_{\rm max}^*-t_i^*\right)$ with ${\bm v}_i^*={\bm P}_i^*/E_i^*$. This gives the relative separation, ${\bm R}_{12}^*= \widetilde{\bm x}_1^*
 -\widetilde{\bm x}_2^*$.
Together with the relative momentum
${\bm K}_{12}^*=({\bm P}_1^*-{\bm P}_2^*)/2$, this defines a
relative phase-space point $({\bm R}_{12}^*,{\bm K}_{12}^*)$ for
each emitted pair, whose ensemble constitutes the classical phase-space distribution,
\begin{equation}
 {\cal S}_{\rm cl}({\bm R}^*,{\bm K}^*)
 =
 \frac{1}{N_{\rm pair}}
 \sum_{\rm pairs}
 \delta^{(3)}({\bm R}^*-{\bm R}_{12}^*)
 \delta^{(3)}({\bm K}^*-{\bm K}_{12}^*).
 \label{eq:Sclassical}
\end{equation} 

Although only the coordinate-space source enters explicitly in the
Koonin--Pratt equation, the source extracted from an event generator
is generally conditional on the relative momentum of the particle
pair. For a given momentum bin, one selects transport pairs according
to their relative momentum ${\bm K}^*$ and constructs the distribution
of their relative separations ${\bm R}^*$. The underlying transport
source is therefore more generally characterized by the joint
phase-space distribution ${\cal S}_{\rm cl}({\bm R}^*,{\bm K}^*)$.
In general, the coordinate and momentum distributions generated by
the transport dynamics need not to factorize,
\begin{equation}
 {\cal S}_{\rm cl}({\bm R}^*,{\bm K}^*)
 \neq
 S_R({\bm R}^*)F_K({\bm K}^*),
 \label{eq:nonfactorization}
\end{equation}
so that the spatial source may depend on the selected relative-momentum
region. Such coordinate--momentum correlations can arise from the
collision geometry, collective expansion, and
particle-production dynamics. 

To incorporate the finite
localization required by quantum mechanics, we associate each
transport phase-space point with a minimum-uncertainty phase-space
distribution. Since the two-body scattering problem entering the
Koonin--Pratt equation is formulated in the PRF, we construct this
distribution in the same frame. For particle $i$, centered at $({\bm R}_i^*,{\bm P}_i^*)$, we adopt
the Gaussian Wigner distribution
\begin{equation}
 W_i({\bm x}^*,{\bm p}^*)
 =N\exp\left[
 -\frac{({\bm x}^*-{\bm R}_i^*)^2}{2\sigma_{x,i}^2}
 -\frac{({\bm p}^*-{\bm P}_i^*)^2}{2\sigma_{p,i}^2}
 \right],
 \label{eq:wigner_single}
\end{equation}
with coordinate and momentum widths constrained by the
minimum-uncertainty condition
\begin{equation}
 \sigma_{x,i}\sigma_{p,i}=\frac{\hbar}{2}.
 \label{eq:uncertainty}
\end{equation}
$N$ is normalization factor. Thus, each classical phase-space point is replaced by a Gaussian phase-space distribution that respects the uncertainty bound.

For the particle pair, we denote the relative variables in the PRF by lowercase letters, ${\bm r}^*={\bm x}^*_1-{\bm x}^*_2$, and ${\bm k}^*=({\bm p}^*_1-{\bm p}^*_2)/2$, while ${\bm R}^*$ and ${\bm K}^*$, defined above, specify the centers of the corresponding relative wave packet for each transport pair. Here, ${\bm r}^*$ and ${\bm k}^*$ are continuous relative phase-space variables within the finite wave packet. The widths of the relative-coordinate and relative-momentum
distributions are $\sigma_r^2=\sigma_{x,1}^2+\sigma_{x,2}^2$ and $\sigma_k^2=(\sigma_{p,1}^2+\sigma_{p,2}^2)/4$, respectively.
For equal single-particle wave-packet widths,
$\sigma_{x,1}=\sigma_{x,2}\equiv\sigma_x$ and
$\sigma_{p,1}=\sigma_{p,2}\equiv\sigma_p$, one obtains $\sigma_r=\sqrt{2}\sigma_x$ and $\sigma_k=\sigma_p/\sqrt{2}$, 
and therefore $\sigma_r\sigma_k=\hbar/2$.

Each classical particle pair, characterized by the relative
phase-space point $({\bm R}_{12}^*,{\bm K}_{12}^*)$, is therefore
replaced by the finite relative phase-space distribution
\begin{equation}
 W_{\rm rel}
 ({\bm r}^*,{\bm k}^*|{\bm R}_{12}^*,{\bm K}_{12}^*)
 =
 G_r({\bm r}^*-{\bm R}_{12}^*)
 G_k({\bm k}^*-{\bm K}_{12}^*),
 \label{eq:Wrelative}
\end{equation}
where
\begin{eqnarray}
G_r({\bm r}^*-{\bm R}_{12}^*)&=&
 \frac{1}{(2\pi\sigma_r^2)^{3/2}}
 \exp\left[
 -\frac{({\bm r}^*-{\bm R}_{12}^*)^2}{2\sigma_r^2}
 \right],\nonumber\\
G_k({\bm k}^*-{\bm K}_{12}^*)&=&
 \frac{1}{(2\pi\sigma_k^2)^{3/2}}
 \exp\left[
 -\frac{({\bm k}^*-{\bm K}_{12}^*)^2}{2\sigma_k^2}
 \right].
\end{eqnarray}
The finite-wave-packet pair distribution is obtained by convoluting
the classical transport distribution with the relative Wigner kernel,
\begin{eqnarray}
 {\cal S}_{\rm wp}({\bm r}^*,{\bm k}^*)
 &=&
 \int d^3R^*\,d^3K^*\,
 {\cal S}_{\rm cl}({\bm R}^*,{\bm K}^*)\nonumber\\
 &\times&W_{\rm rel}
 ({\bm r}^*,{\bm k}^*|{\bm R}^*,{\bm K}^*).
\end{eqnarray}
Here, $({\bm R}^*,{\bm K}^*)$ are integration variables over the
classical relative phase space, whose discrete values in the transport
simulation are the pair centers
$({\bm R}_{12}^*,{\bm K}_{12}^*)$.

For a given relative momentum ${\bm k}^*$, the normalized emission source entering the Koonin--Pratt formalism~\eqref{eq:KP_intro} is the
conditional distribution, which can be expressed as,
\begin{equation}
 S_{\rm wp}({\bm r}^*|{\bm k}^*)
 =
 \frac{
 {\cal S}_{\rm wp}({\bm r}^*,{\bm k}^*)
 }{
 \int d^3r^*\,
 {\cal S}_{\rm wp}({\bm r}^*,{\bm k}^*)
 }.
 \label{eq:conditional_source}
\end{equation}
The classical emission source ${\cal S}_{\rm cl}$~\eqref{eq:Sclassical} introduced above is normalized analogously at fixed ${\bm k}^*$.

Importantly, ${\bm k}^*$ appearing in the scattering wave function is
the asymptotic relative momentum of the observed pair in its
center-of-mass frame. The momentum-space width $\sigma_k$ should
therefore not be interpreted as an experimental momentum resolution.
Rather, it represents the intrinsic momentum uncertainty associated
with the finite localization of the emitted particles and thereby
modifies the conditional emission source at a given asymptotic
relative momentum. Consequently, finite quantum localization affects
not only the coordinate-space distribution of the emission points but,
in the presence of coordinate--momentum correlations, also the
coordinate source selected at a given relative momentum.

\begin{figure}[!htb]
    \centering    \includegraphics[width=0.45\textwidth]{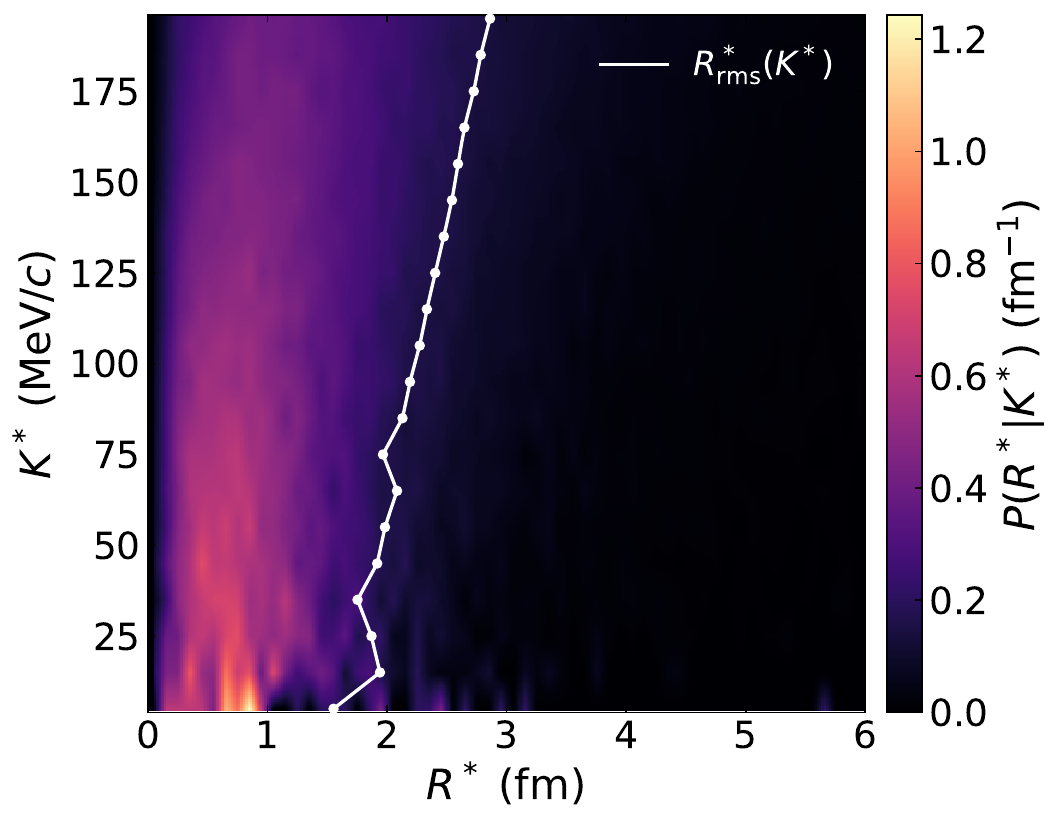}   
   \caption{Conditional relative-distance distribution $P(R^*|K^*)$ of proton
pairs obtained from PHSD simulations of $pp$ collisions at
$\sqrt{s}=7~{\rm TeV}$. The white curve and symbols show the
root-mean-square separation $R_{\rm rms}^*(K^*)$.}
    \label{fig.1}
\end{figure}

We first characterize the classical phase-space distribution.
To quantify its coordinate--momentum correlations, we introduce the
conditional relative-distance distribution
\begin{equation}
P(R^*|K^*)=
\frac{d^2N/(dR^*dK^*)}{dN/dK^*},
\label{eq:conditional_source}
\end{equation}
which is normalized for each $K^*$ and gives the probability density of finding a pair with relative
separation $R^*=|{\bm R}^*|$ at a given relative momentum
$K^*=|{\bm K}^*|$. Figure~\ref{fig.1} shows the conditional relative-distance
distribution $P(R^*|K^*)$ of proton pairs obtained from PHSD
simulations of $pp$ collisions at $\sqrt{s}=7~{\rm TeV}$.
In high energy $pp$ collisions, proton production in PHSD is initialized through Lund-string excitation and fragmentation, implemented using PYTHIA/FRITIOF, followed by the dynamical propagation, resonance decays, and possible final-state hadronic interactions~\cite{Kireyeu:2020wou}.
A pronounced dependence of the source on the pair relative momentum
is observed. At small $K^*$, the distribution is concentrated at
relatively small separations, whereas with increasing $K^*$ it
gradually shifts toward larger $R^*$ and becomes broader.
This trend is also reflected in the root-mean-square separation
$R_{\rm rms}^*(K^*)\equiv
\sqrt{\langle(R^*)^2\rangle}$, shown by the white curve.
The observed $K^*$ dependence demonstrates that the relative
coordinate and momentum of the emitted pairs are correlated, i.e.,
the classical pair phase-space distribution cannot in general be
factorized as
${\cal S}_{\rm cl}({\bm R}^*,{\bm K}^*)
\neq S_R({\bm R}^*)F_K({\bm K}^*)$.
Such coordinate--momentum correlations are essential in the
wave-packet formulation, since the finite momentum width mixes
neighboring $K^*$ regions with different spatial distributions.

\begin{figure}[!htb]
    \centering    \includegraphics[width=0.45\textwidth]{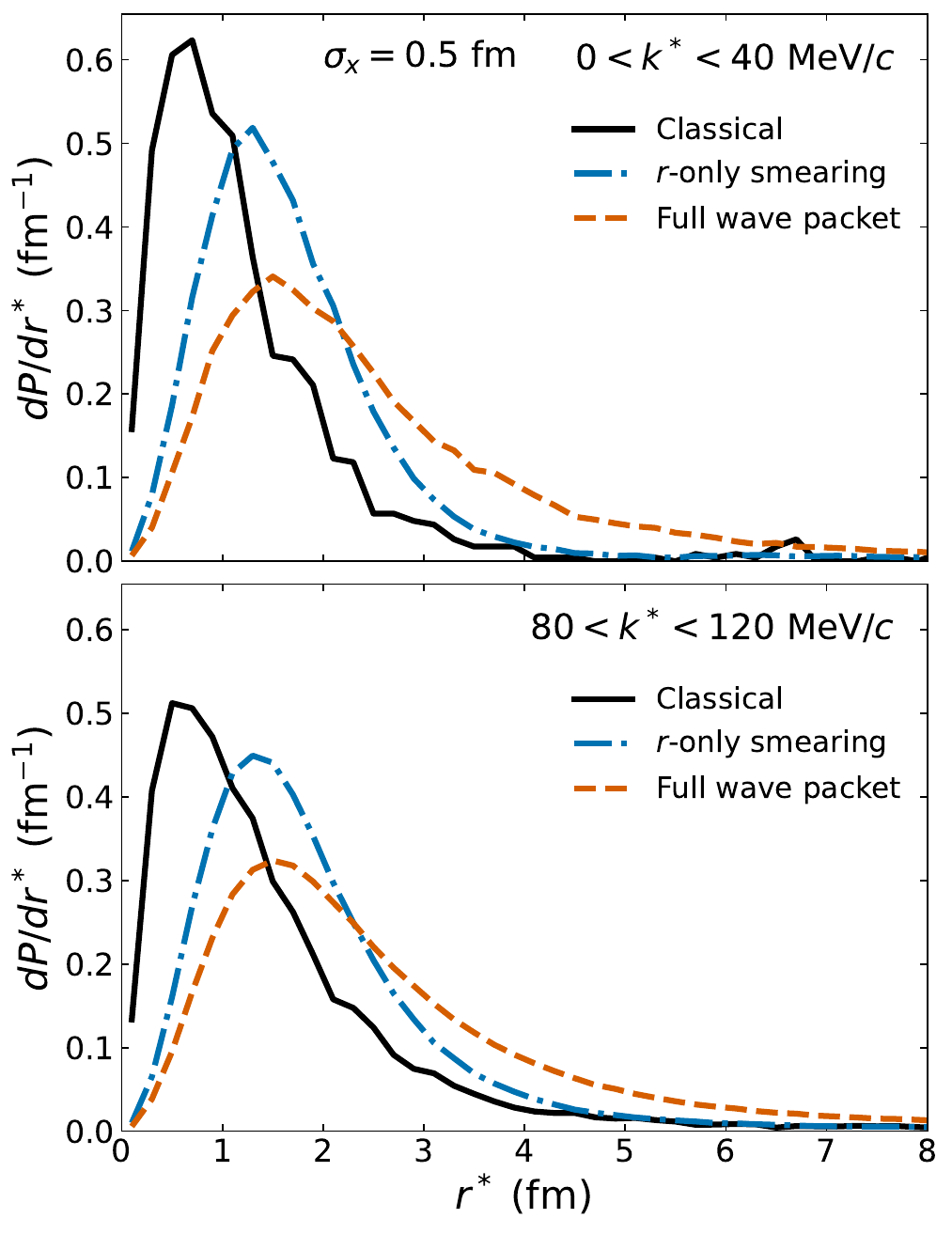}   
   \caption{Comparison of the classical source (solid lines) and the source after finite wave-packet smearing in coordinate space (blue-dash-dotted lines) and full phase-space (orange-dashed lines) for $0<k^*<40$ MeV/c (upper panel) and $80<k^*<120$ MeV/c (lower panel). A single-particle spatial width of $\sigma_x=$ 0.5 fm is employed in this plot.}
    \label{fig.2}
\end{figure}
Since the spatial localization width of a hadron produced in a transport calculation is not uniquely determined, we treat $\sigma_x$ as the only model parameter. The effect of finite quantum localization on the emission source is illustrated in Fig.~\ref{fig.2} for two relative-momentum intervals. Coordinate-space smearing alone (orange-dashed lines) broadens
the source and shifts its strength toward larger $r^*$. The full
wave-packet treatment (blue-dash-dotted lines) leads to an additional modification because the
finite momentum width mixes neighboring $K^*$ regions with different
spatial distributions. The clear difference between the $r$-only and
full wave-packet results therefore shows that quantum localization
cannot, in general, be reduced to a simple increase of the source
radius, but probes the nonfactorizable phase-space structure of the
transport source.

Now, let's investigate the influence on the particle femtoscopic correlations. The proton--proton scattering wave function entering the
Koonin--Pratt equation is calculated using the\textit{Argonne}-18 potential~\cite{Wiringa:1994wb}
together the Coulomb potential, with the two-proton wave function properly antisymmetrized. For
unpolarized protons, the correlation is averaged over the spin-singlet
and spin-triplet channels with statistical weights $1/4$ and $3/4$,
respectively. Moreover, the measured
$pp$ correlation contains both primordial and feed-down contributions,
which are combined according to their corresponding pair fractions,
\begin{equation}
C_{pp}(k^*)=
f_{\rm pri} C_{pp}(k^*) + \sum_Y f_{Y \to p} C_{pp}^{pY\rightarrow pp}(k^*),
\end{equation}
where $f_{\rm pri}$ and $f_{Y\to p}$ denote the fractions of primordial produced and those originating from hyperon decays, respectively. The previous experimental study shows $f_{\rm pri}=74.18\%$ and the most dominated hyperon decay is $f_{\Lambda\to p}=15.51\%$~\cite{ALICE:2018ysd}. The residual $pp$ correlation generated from the parent $p\Lambda$ correlation $C_{pp}^{p\Lambda\rightarrow pp}$ is calculated and shown in Appendix~\ref{sec.appendix}.
\begin{figure}[!htb]
    \centering    \includegraphics[width=0.45\textwidth]{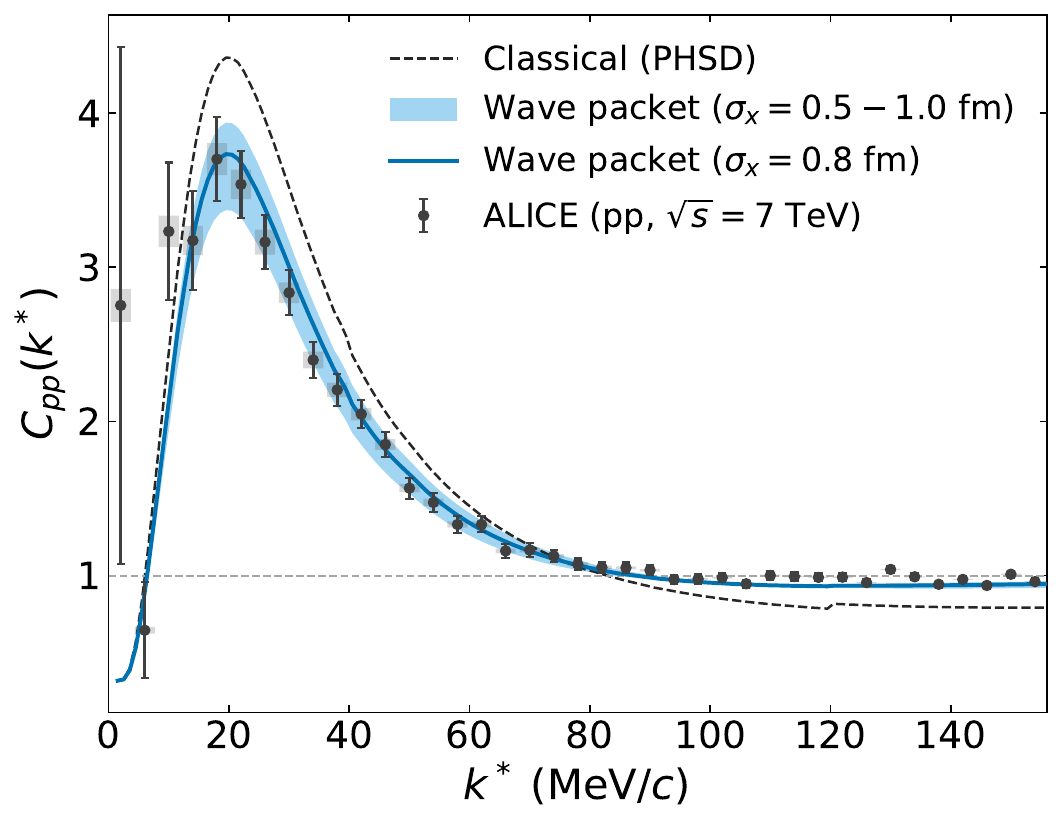}   
   \caption{Proton-proton correlation functions calculated using the classical source (black line) and the finite wave-packet source with $\sigma_x=0.8~\rm fm$ (blue line). The light-blue band is with $\sigma_x=0.5-1.0~\rm fm$, compared with experimental data, which is from $pp$ collisions at $\sqrt{s}=7~\rm TeV$ (dots)~\cite{ALICE:2018ysd}.}
    \label{fig.corrpp}
\end{figure}

Since femtoscopic correlations are particularly sensitive to the source distribution at small relative separations, such modifications can have a sizable impact on the calculated $pp$ correlation function. 
The impact of finite quantum localization on the observable correlation function is shown in Fig.~\ref{fig.corrpp}. We calculate the $pp$ correlation using the same two-proton scattering wave function, while changing only the emission source from the classical PHSD distribution to its wave-packet representation. A sizable modification is observed. The classical source produces a pronounced correlation peak, whereas the finite wave-packet source substantially reduces its magnitude. This reduction originates from the redistribution of source probability from small to larger relative separations shown in Fig.~\ref{fig.2}, to which the short-range part of the femtoscopic kernel is particularly sensitive. With $\sigma_x=0.8~\rm fm$, the observed correlation function can be well described with emission source composed of wave-packets.

The wave-packet formulation also provides a natural criterion for the validity of the conventional point-emitter approximation. Finite quantum localization introduces two characteristic scales:
the relative-coordinate width $\sigma_r$ and the relative-momentum
width $\sigma_k$. Their importance should be compared with the
corresponding scales over which the underlying emission function
varies. We therefore introduce the dimensionless quantities
\begin{equation}
    \epsilon_r =
    \frac{\sigma_r}{R_{\rm source}},
    \qquad
    \epsilon_k =
    \frac{\sigma_k}{\Delta K},
    \label{eq:epsilon}
\end{equation}
where $R_{\rm source}$ characterizes the spatial extent of the
emission source, while $\Delta K$ denotes the characteristic
relative-momentum scale over which the transport phase-space
distribution changes appreciably.

The conventional point-emitter description is expected to be
insensitive to finite quantum localization when both $\epsilon_r \ll 1$ and $\epsilon_k \ll 1$.
The first condition ensures that the coordinate-space smearing is
small compared with the spatial extent of the source. The second
ensures that the momentum uncertainty does not significantly mix
regions of the transport phase-space distribution characterized by
different relative momenta and, consequently, different conditional
spatial sources.

These two conditions, however, cannot be satisfied independently.
For the relative wave packet, the coordinate and momentum widths
obey the uncertainty relation $\sigma_r \sigma_k \geq \hbar/2$. Consequently, $\epsilon_r\epsilon_k\geq \hbar/(2R_{\rm source}\Delta K)$. For the minimum-uncertainty Gaussian wave packets considered in this study, the equality is realized. A regime in which both finite-localization corrections are
simultaneously small therefore requires
\begin{equation}
    R_{\rm source}\Delta K
    \gg \frac{\hbar}{2}.
    \label{eq:semiclassical_condition}
\end{equation}
This relation provides a simple phase-space criterion for the
applicability of the classical point-emitter approximation in
event-generator-based femtoscopy. When the characteristic source
size becomes comparable to the localization width, or when the
emission function varies rapidly with relative momentum, finite
wave-packet effects can no longer be assumed to be negligible.

For small femtoscopic sources, $R_{\rm source}$ is typically of
order $1$~fm~\cite{ALICE:2023sjd,ALICE:2020ibs,Wang:2024bpl}, while the relevant momentum dependence of the source
may occur over scales of order $100$~MeV/$c$. The product
$R_{\rm source}\Delta K$ is then not parametrically larger than
$\hbar/2$. This indicates that small collision systems naturally
approach a regime in which the point-emitter approximation should
be quantitatively tested rather than assumed. The PHSD results
presented above provide an explicit example: the observed
coordinate--momentum correlations make the finite momentum width
of the wave packet relevant for the conditional source
$S(\bm r^*|\bm k^*)$ and, consequently, for the resulting
femtoscopic correlation function.

In summary, we have demonstrated that the finite quantum localization
of emitted particles constitutes an important ingredient of
transport-based femtoscopy that is absent in the conventional
point-emitter construction of the emission source. By incorporating
the minimum phase-space uncertainty through a Wigner formulation, we
show that quantum localization affects both the relative coordinate
and momentum distributions. In realistic transport sources,
coordinate--momentum correlations couple the momentum uncertainty to
the spatial distribution, leading to a nontrivial,
momentum-dependent modification of the emission source.

Using PHSD simulations of $pp$ collisions at
$\sqrt{s}=7~\mathrm{TeV}$, we demonstrate that this quantum effect
produces sizable modifications of the proton-pair emission source and,
consequently, of the low-$k^*$ femtoscopic correlation. Quantum
localization is therefore not merely a formal correction to the
classical source construction, but can directly affect femtoscopic
observables and the extraction of hadron--hadron interactions.

More generally, the uncertainty relation yields the phase-space
criterion $R_{\rm source}\Delta K \gg \hbar/2$
for the classical point-emitter approximation to be valid. Small
collision systems naturally approach the regime in which this
condition is no longer parametrically satisfied. Our results therefore
identify finite quantum localization as an intrinsic limitation of the
classical point-emitter construction and establish when this quantum
effect must be accounted for in precision femtoscopy.

\textbf{Acknowledgment.--}
We acknowledge the support by the Deutsche Forschungsgemeinschaft (DFG, German Research Foundation) through the grant CRC-TR 211 'Strong-interaction matter under extreme conditions' - Project number 315477589 - TRR 211. The computational resources have been provided by the Center for Scientific Computing (CSC) of the Goethe University, Frankfurt.

\bibliographystyle{apsrev4-1}
\bibliography{refs}

\appendix
\section{Feed-down contribution from $\Lambda\rightarrow p\pi^-$}
\label{sec.appendix}

The experimentally measured proton--proton correlation may contain
residual correlations originating from correlated $p\Lambda$ pairs,
where the $\Lambda$ subsequently undergoes the weak decay
$\Lambda\rightarrow p\pi^-$~\cite{ParticleDataGroup:2022pth}. The correlation of the parent
$p\Lambda$ pair is then transferred to the daughter $pp$ pair through
the decay kinematics. We account for this effect using a decay
response matrix constructed by Monte Carlo simulation.

For a parent $p\Lambda$ pair with relative momentum
$k_{p\Lambda}^*$, we first work in the pair rest frame (PRF), where
the four-momenta can be written as
\begin{equation}
 p_p^\mu =
 \left(E_p,0,0,-k_{p\Lambda}^*\right),
 \qquad
 p_\Lambda^\mu = \left(E_\Lambda,0,0,+k_{p\Lambda}^*\right),
 \label{eq:plambda_prf}
\end{equation}
with
\begin{equation}
 E_p=\sqrt{m_p^2+(k_{p\Lambda}^*)^2},
 \qquad
 E_\Lambda=\sqrt{m_\Lambda^2+(k_{p\Lambda}^*)^2}.
\end{equation}
The choice of the $z$ axis in Eq.~(\ref{eq:plambda_prf}) is arbitrary
because only the magnitude of the relative momentum is relevant.

The decay $\Lambda\rightarrow p\pi^-$ is generated in the
$\Lambda$ rest frame. For the two-body decay, the magnitude of the
daughter momentum is fixed by energy--momentum conservation,
\begin{equation}
 q_{\Lambda\rightarrow p\pi}
 =
 \frac{
 \sqrt{\lambda(m_\Lambda^2,m_p^2,m_\pi^2)}
 }{2m_\Lambda},
 \label{eq:lambda_decay_q}
\end{equation}
where $ \lambda(x,y,z)=x^2+y^2+z^2-2xy-2xz-2yz$ is the K\"all\'en function. Numerically,
$q_{\Lambda\rightarrow p\pi}\simeq100.6$~MeV/$c$.
For an unpolarized $\Lambda$, the decay direction is sampled
isotropically,
\begin{equation}
 \frac{dP}{d\Omega}=\frac{1}{4\pi},
\end{equation}
such that the daughter-proton four-momentum in the $\Lambda$ rest
frame is
\begin{equation}
 p_{p_d}^{\mu}
 =
 \left(
 \sqrt{m_p^2+q^2},
 q\sin\theta\cos\phi,
 q\sin\theta\sin\phi,
 q\cos\theta
 \right).
 \label{eq:daughter_p}
\end{equation}

The daughter proton is subsequently boosted from the $\Lambda$ rest
frame to the original $p\Lambda$ PRF. Together with the primary
proton, it defines the daughter $pp$ pair. Since the rest frame of
the daughter $pp$ pair is in general different from the parent
$p\Lambda$ PRF, the two proton four-momenta are further transformed
to their own pair rest frame. The resulting daughter relative
momentum is
\begin{equation}
 k_{pp}^*
 =
 \frac{1}{2}
 \left|
 \mathbf p_{p_1}^{\,*}
 -
 \mathbf p_{p_2}^{\,*}
 \right|
 =
 \left|\mathbf p_{p_1}^{\,*}\right|
 =
 \left|\mathbf p_{p_2}^{\,*}\right|.
 \label{eq:kpp_decay}
\end{equation}

\begin{figure}[!htb]
    \centering
    \includegraphics[width=0.45\textwidth]{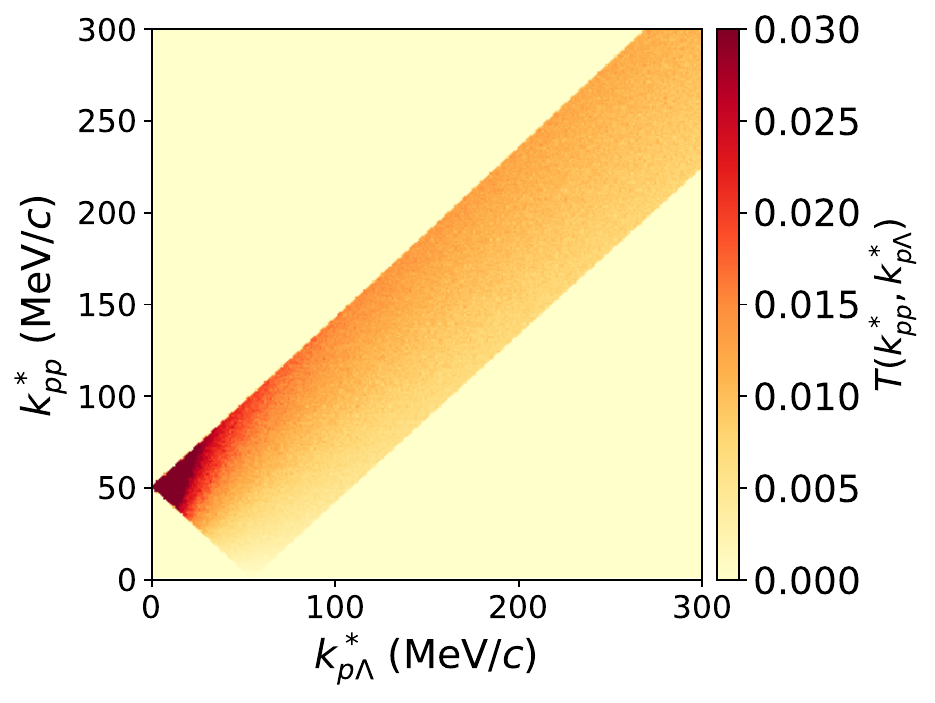} 
    \caption{The response matrix $T_{ji}$ for feed-down contributions from $\Lambda$ to proton.}
    \label{fig.Tmatrix}
\end{figure}
Repeating the decay simulation for each parent relative-momentum
bin defines the response matrix
\begin{equation}
 T_{ji}
 \equiv
 T(k_{pp,j}^*,k_{p\Lambda,i}^*)
 =
 P\left(
 k_{pp}^*\in j
 \,\big|\,
 k_{p\Lambda}^*\in i
 \right),
 \label{eq:decay_matrix}
\end{equation}
which gives the probability that a parent $p\Lambda$ pair in
momentum bin $i$ contributes to a daughter $pp$ pair in momentum
bin $j$. The matrix is normalized for each parent momentum bin, $\sum_j T_{ji}=1$. The response matrix therefore contains only the kinematic
redistribution of the relative momentum induced by the
$\Lambda\rightarrow p\pi^-$ decay. The response matrix is calculated numerically and shown in Fig.~\ref{fig.Tmatrix}.

The residual $pp$ correlation generated from the parent
$p\Lambda$ correlation is obtained by folding
$C_{p\Lambda}(k_{p\Lambda}^*)$ with the decay response matrix.
In the approximation that the uncorrelated parent-pair spectrum
varies slowly over the considered momentum range, this gives
\begin{equation}
 C_{pp}^{p\Lambda\rightarrow pp}(k_{pp,j}^*)
 =
 \frac{\sum_i
 T_{ji}\,
 C_{p\Lambda}(k_{p\Lambda,i}^*)
 }{\sum_i T_{ji}
 }.
 \label{eq:plambda_pp_feeddown}
\end{equation}
This procedure preserves the correlation carried by the parent
$p\Lambda$ pair while accounting for the redistribution and
smearing of relative momentum caused by the weak decay of the
$\Lambda$. 
\begin{figure}[!htb]
    \centering
    \includegraphics[width=0.45\textwidth]{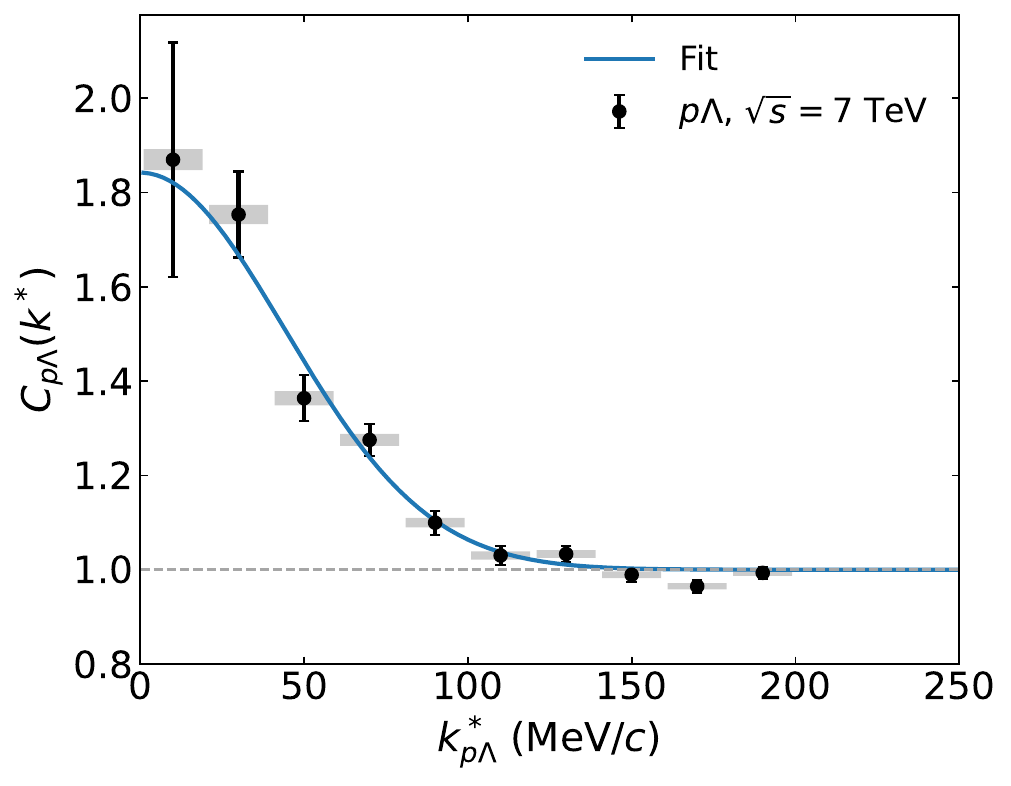} 
    \includegraphics[width=0.45\textwidth]{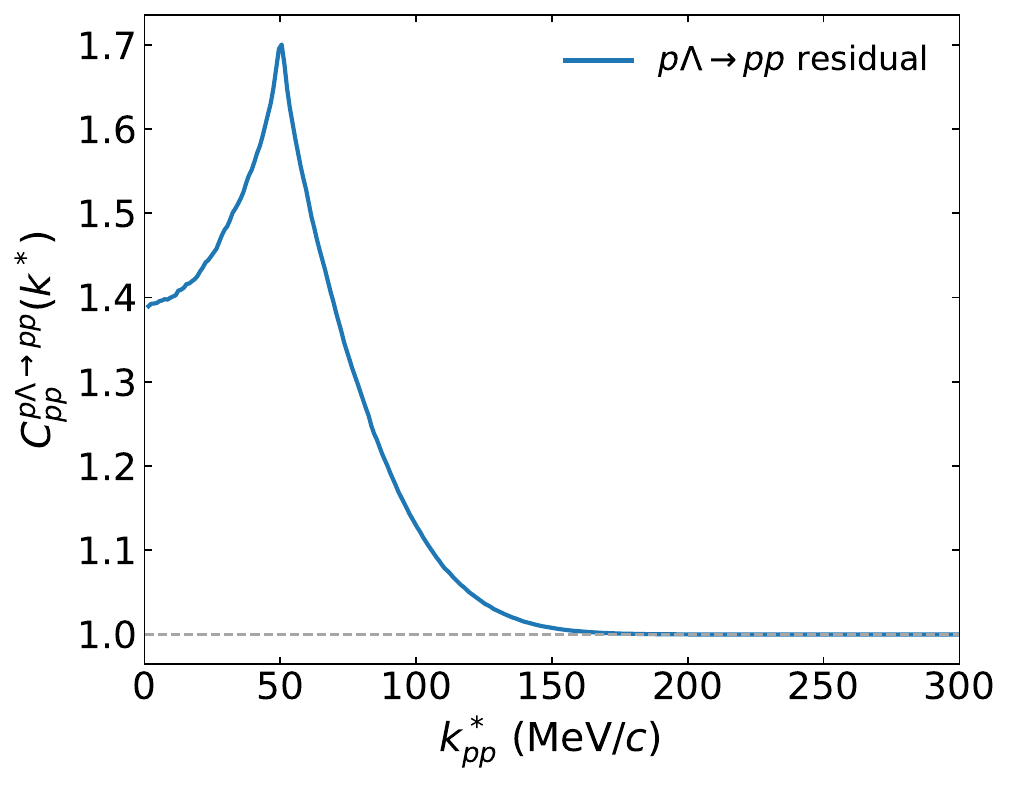} 
    \caption{The $p\Lambda$ correlation (top panel) from the experiment at $\sqrt{s}=7~\rm TeV$~\cite{ALICE:2018ysd},
    and residual $pp$ correlation (bottom panel) calculated by Eq.~\eqref{eq:plambda_pp_feeddown}.}
    \label{fig.platopp}
\end{figure}

The experimentally measured $p\Lambda$ correlation function and its
parametrization are shown in the top panel of Fig.~\ref{fig.platopp}. Since the
experimental correlation is available only at a finite number of
relative-momentum points, while the decay transformation requires the
parent correlation over a continuous range of $k^*_{p\Lambda}$, we
parametrize the measured correlation by the simple form
\begin{equation}
 C_{p\Lambda}^{\rm fit}(k^*)
 =
 1+A\exp\left[-\left(\frac{k^*}{\kappa}\right)^2\right].
 \label{eq:pLambda_fit}
\end{equation}
The fit parameters are found to be
$A=0.842$, $\kappa=62.3~{\rm MeV}/c$,
with $\chi^2/{\rm ndf}=1.52$. This parametrization provides a smooth
description of the measured low-$k^*$ enhancement and naturally
approaches unity at large relative momentum. The latter property is
particularly useful for the decay transformation, since parent
$p\Lambda$ pairs outside the experimentally measured momentum range
can contribute to daughter $pp$ pairs at lower relative momentum.

The resulting residual correlation is shown in the bottom panel of 
Fig.~\ref{fig.platopp}. Although the original $p\Lambda$
correlation has its maximum at $k^*_{p\Lambda}\simeq0$, the daughter
$pp$ correlation develops a pronounced maximum around
$k^*_{pp}\simeq50$~MeV/$c$. This shift is a consequence of the decay
kinematics rather than a new interaction between the two daughter
protons. In the $\Lambda$ rest frame, the proton from the two-body
decay carries a fixed momentum
\begin{equation}
 q_{\Lambda\rightarrow p\pi}
 =
 \frac{
 \sqrt{\lambda(m_\Lambda^2,m_p^2,m_\pi^2)}
 }{2m_\Lambda}
 \simeq100.6~{\rm MeV}/c.
 \label{eq:lambda_decay_q}
\end{equation}
For the strongly correlated parent pairs at small
$k^*_{p\Lambda}$, the primary proton and the $\Lambda$ have only a
small relative momentum. The decay therefore introduces a momentum
difference predominantly determined by
$q_{\Lambda\rightarrow p\pi}$. Since the relative momentum of the
daughter $pp$ pair is defined in its pair rest frame as
\begin{equation}
 k^*_{pp}
 =
 \frac{1}{2}
 \left|
 \mathbf p^{\,*}_{p,1}
 -
 \mathbf p^{\,*}_{p,2}
 \right|,
\end{equation}
the characteristic decay momentum translates approximately into a
daughter relative-momentum scale
\begin{equation}
 k^*_{pp}
 \sim
 \frac{q_{\Lambda\rightarrow p\pi}}{2}
 \sim50~{\rm MeV}/c.
\end{equation}
The exact transformation is not one-to-one: the Lorentz boost of the
decay proton, the finite parent relative momentum, and the isotropic
decay angle generate a distribution of daughter momenta. These
effects are encoded in the response matrix $T_{ji}$ and broaden the
kinematic structure around this characteristic scale. Consequently,
the peak near $50$~MeV/$c$ in the residual $pp$ correlation should
be interpreted as the decay image of the strong low-$k^*$
$p\Lambda$ correlation.

\end{document}